\documentclass[]{spie}  %>>> use for US letter paper
\usepackage{amsmath,amsfonts,amssymb}
\usepackage{graphicx}
 \usepackage{booktabs} 
 \usepackage{float}
\usepackage{balance}
\usepackage{multirow} 
\usepackage[colorlinks=true, allcolors=blue]{hyperref}
\usepackage{eso-pic}

\title{Quantitative Measures for Passive Sonar Texture Analysis}

\author[a]{Jarin Ritu}
\author[b]{Alexandra Van Dine}
\author[a]{Joshua Peeples}

\affil[a]{Department of Electrical and Computer Engineering, Texas A\&M University, College Station, TX, USA}
\affil[b]{Massachusetts Institute of Technology Lincoln Laboratory, Lexington, MA, USA}

\authorinfo{Further author information: (Send correspondence to J.P.)\\J.P.: E-mail: jpeeples@tamu.edu, Telephone: +1 979 458-7026}

\begin{document} 
\maketitle

\begin{abstract}
Passive sonar signals contain complex characteristics often arising from environmental noise, vessel machinery, and propagation effects. While convolutional neural networks (CNNs) perform well on passive sonar classification tasks, they can struggle with statistical variations that occur in the data. To investigate this limitation,  synthetic underwater acoustic datasets are generated that centered on amplitude and period variations. Two metrics are proposed to quantify and validate these characteristics in the context of statistical and structural texture for passive sonar. These measures are applied to real-world passive sonar datasets to assess texture information in the signals and correlate the performances of the models. Results show that CNNs underperform on statistically textured signals, but incorporating explicit statistical texture modeling yields consistent improvements. These findings highlight the importance of quantifying texture information for passive sonar classification. The code is publicly available at: \url{https://github.com/Advanced-Vision-and-Learning-Lab/Passive_Sonar_Texture_Analysis}.

\end{abstract}
\footnotetext{DISTRIBUTION STATEMENT A. Approved for public release. Distribution is unlimited. This material is based upon work supported by the Under Secretary of War for Research and Engineering under Air Force Contract No. FA8702-15-D-0001 or FA8702-25-D-B002. Any opinions, findings, conclusions or recommendations expressed in this material are those of the author(s) and do not necessarily reflect the views of the Under Secretary of Defense for Research and Engineering. \textsuperscript{\textcopyright} 2025 Massachusetts Institute of Technology. Delivered to the U.S. Government with Unlimited Rights, as defined in DFARS Part 252.227-7013 or 7014 (Feb 2014). Notwithstanding any copyright notice, U.S. Government rights in this work are defined by DFARS 252.227-7013 or DFARS 252.227-7014 as detailed above. Use of this work other than as specifically authorized by the U.S. Government may violate any copyrights that exist in this work.}

\section{Introduction}
\label{sec:intro}

Texture is a fundamental property in both image and audio domains, capturing the underlying structural and statistical patterns in signals \cite{LIM}. In images, texture refers to spatial elements such as edges, repetitive structures, and intensity variations. In the audio domain, structural texture corresponds to periodic features like harmonics and modulations \cite{sharma2022trends}, while statistical texture reflects stochastic variations in amplitude and frequency \cite{trevorrow2021examination}. These characteristics influence how models interpret signals, especially for complex tasks like passive sonar classification. Variations in passive sonar arise from environmental noise, turbulence, and reflections from dynamic sources \cite{neupane2020review}. While conventional features such as MFCCs, Gammatone filters, and wavelet transforms are widely used, they often fail to capture how statistical randomness interacts with structured periodicity \cite{abdul2022mel, lian2017underwater}.

Prior work has explored statistical descriptors including mean, variance, skewness, entropy, and other higher-order measures \cite{gurban2008using, peeples2021histogram} to distinguish between noise-like and harmonically rich signals \cite{barchiesi2015acoustic}. Structural texture has been examined using methods such as autocorrelation and harmonic envelope modeling, which capture temporal repetition and spectral regularity \cite{starovoitov1998texture}. Despite their importance, most deep models like CNNs are primarily optimized for local structural patterns \cite{peeples2021histogram}, and often underperform in capturing statistical texture \cite{peeples2022histogram, ji2022structural}.

The Histogram Layer Time-Delay Neural Network (HLTDNN) \cite{ritu2023histogram} demonstrated that incorporating statistical representations improves passive sonar classification. Building on this, this work investigates whether explicitly quantifying audio texture can better guide model selection. Synthetic datasets are developed to validate two texture quantification metrics: one entropy-based for statistical texture and another autocorrelation-based for structural texture. These are also applied to real datasets to analyze how texture composition relates to performance.

\vspace{0.5em}
\noindent The main contributions of this work are as follows:
\begin{itemize}
    \item Synthetic underwater acoustic dataset with controlled statistical and structural textures variations.
    \item Formal definitions to quantify passive sonar texture.
    \item Empirical analysis demonstrating the correlation between texture characteristics of the data and model performance.
\end{itemize}

\section{Methodology}
Three synthetic datasets were created to simulate sonar-like recordings with controlled texture characteristics: (1) statistical texture based on amplitude variation, (2) structural texture based on periodic patterns, and (3) mixed statistical and structural textures. Each dataset contains four vessel classes with 10,000 samples per class, totaling 40,000 samples. The design was informed by patterns observed in the DeepShip dataset~\cite{irfan2021deepship}, where cargo and tanker vessels show stronger low-frequency signatures, and passengership and tug vessels show more high-frequency activity. To reflect this, the synthetic datasets assign lower frequencies to Classes~1 and~2 and higher frequencies to Classes~3 and~4. Spectral peaks were visually inspected using STFT plots of DeepShip recordings, where bright, stable regions in the frequency axis were noted across multiple samples per class. Larger vessels consistently exhibited high-energy content between 2.5--4.5\,kHz, while smaller vessels showed dominant peaks between 5.0--6.5\,kHz, guiding the selection of class-wise base frequencies.

\begin{figure*}[t]
\centering
\includegraphics[width=1\linewidth]{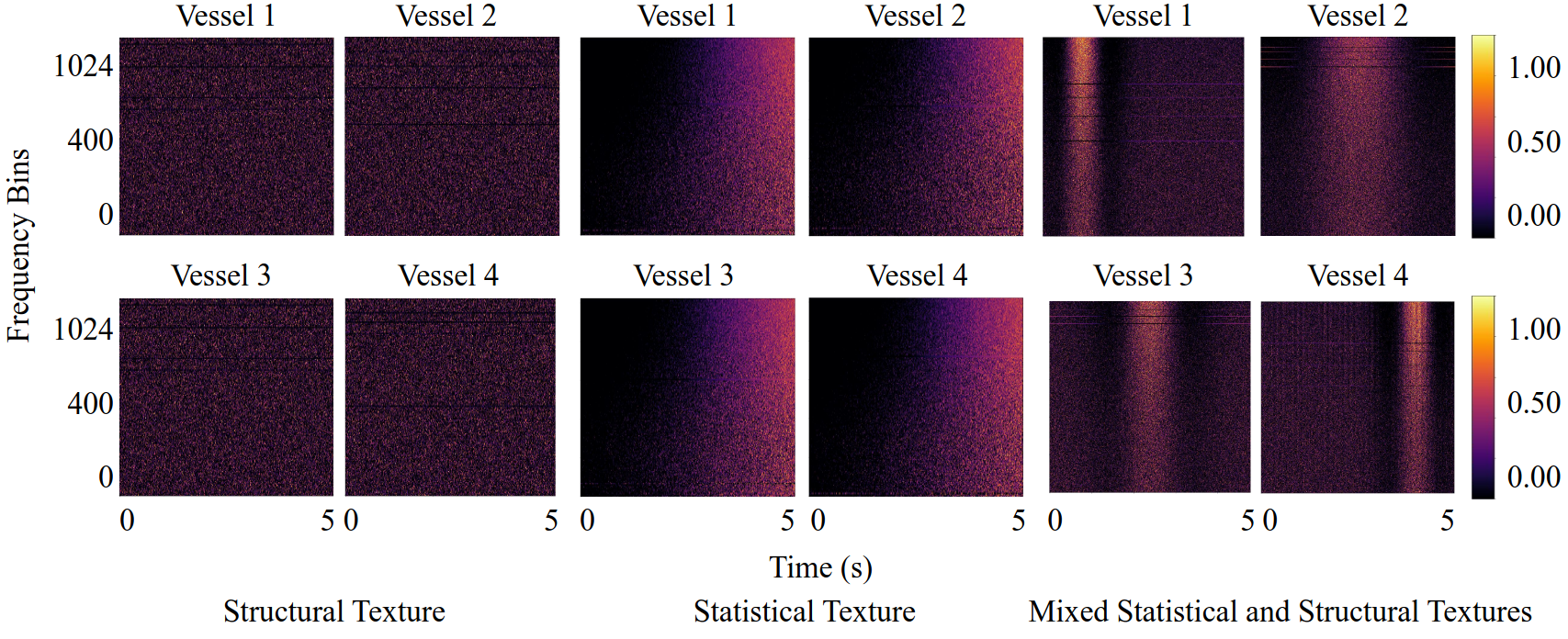}
\caption{Example spectrograms showing one sample per class (Vessel 1, Vessel 2, Vessel 3, Vessel 4) for each synthetic dataset type: Structural Texture, Statistical Texture, and Mixed Statistical and Structural Textures.}
\label{figure:dataset_examples}
\end{figure*}

Amplitude distributions were modeled using Rayleigh and K-distributions, common in sonar signal modeling~\cite{trevorrow2021examination}, with parameters empirically chosen to provide diverse yet realistic signal textures. Amplitude modulation depth and rate were sampled from ranges of [0.1, 0.3] and [1, 3]~Hz, and noise levels from \([0.006, 0.01]\), ensuring that background effects do not overwhelm key signal characteristics. As summarized in Table~\ref{tab:balanced_dataset}, each vessel class is configured with distinct frequency, amplitude, and modulation properties. Figure~\ref{figure:dataset_examples} shows representative spectrograms from each dataset. The full dataset generation procedures are described in Sections~\ref{sect:stats} through~\ref{sect:mixed}.

\begin{table}[ht]
\caption{Class-wise configuration for the mixed statistical and structural textures dataset. Frequencies are expressed in kilohertz (kHz).}
\label{tab:balanced_dataset}
\begin{center}
\small
\setlength{\tabcolsep}{8pt}
\renewcommand{\arraystretch}{1.2}
\begin{tabular}{|l|c|c|c|}
\hline
\textbf{Vessel Class} & \textbf{Frequencies (kHz)} & \textbf{Amplitude Distributions} & \textbf{Modulation Rate} \\
\hline
Vessel 1 & \{2.2, 3.5, 5.0, 6.5\} & K $\rightarrow$ Rayleigh & Slow \\ \hline
Vessel 2 & \{3.0, 4.5, 6.0, 7.0\} & Rayleigh $\rightarrow$ K & Slow \\ \hline
Vessel 3 & \{9.0, 10.5, 12.0, 13.0\} & Rayleigh $\rightarrow$ K & Fast \\ \hline
Vessel 4 & \{10.0, 11.5, 13.0, 14.0\} & K $\rightarrow$ Rayleigh & Fast \\
\hline
\end{tabular}
\end{center}
\end{table}

\subsection{Statistical Texture Dataset}
\label{sect:stats}

This dataset emphasizes stochastic variation in amplitude and background noise, with partial spectral overlap between adjacent classes. Each signal contains three frequency components per class. The amplitude envelope \( A^f(t) \), defined for each frequency \( f \), is sampled as:

\begin{equation}
A^f(t) \sim 
\begin{cases}
\text{Rayleigh}(\sigma), & \text{with probability } p, \\
\text{K}(k, \theta), & \text{with probability } 1 - p
\end{cases}
\label{eqn:amp}
\end{equation}

To ensure smooth transitions, sinusoidal amplitude modulation is applied segment-wise. Background noise \( \eta(t) \) is added as:

\begin{equation}
\eta(t) = \nu(t) + \sum_{k \in \mathcal{I}} a_k \cdot \delta(t - t_k),
\label{eqn:background}
\end{equation}

\noindent where \( \nu(t) \) is colored noise and impulses \( a_k \) simulate short acoustic bursts. The design maintains amplitude-driven variation while limiting periodic structure.

\subsection{Structural Texture Dataset}
\label{sect:struct}

This dataset emphasizes deterministic temporal structure with minimal randomness. Each signal consists of three harmonics derived from a class-specific base frequency \( f_c \), perturbed by uniformly sampled jitter:

\begin{equation}
f_k = k \cdot f_c + \delta_k, \quad \delta_k \sim \mathcal{U}[-50, 50],
\label{eqn:harmonics}
\end{equation}

Each class uses a distinct amplitude envelope (e.g., triangular, exponential decay) defined over normalized time. Low-amplitude Gaussian noise is added to preserve harmonic clarity while introducing mild background variation.

\subsection{Mixed Statistical and Structural Textures Dataset}
\label{sect:mixed}

This dataset simulates acoustic signals exhibiting both amplitude variation and periodic structure, capturing realistic underwater texture characteristics where random fluctuations and deterministic patterns coexist. Each amplitude envelope \( \hat{A}^f(t) \) is generated by modulating a probabilistic envelope with a sinusoidal component:

\begin{equation}
\hat{A}^f(t) = \alpha(t) \cdot \left(1 + d \cdot \sin(2\pi r t)\right),
\label{eqn:amplitude_2}
\end{equation}

\noindent where \( \alpha(t) \sim \text{Rayleigh}(\sigma) \) or \( \text{K}(k, \theta) \), and \( d \), \( r \) are class-specific modulation depth and rate parameters.

To enable smooth transitions between amplitude distributions, a Gaussian-based blending function interpolates between Rayleigh and K-distributed envelopes:

\begin{equation}
\alpha(t) = w(t) \cdot \alpha_{\text{Rayleigh}}(t) + \left(1 - w(t)\right) \cdot \alpha_{\text{K}}(t),
\label{eqn:weight}
\end{equation}

\noindent where \( w(t) = \exp\left(-\frac{(t - \mu)^2}{2\sigma^2}\right) \), with \( \mu \) and \( \sigma \) controlling the transition center and width.

Each class employs distinct frequency bands, amplitude transition directions (e.g., K~\(\rightarrow\)~Rayleigh), and modulation speeds. Lower-frequency classes (Vessel 1 and 2) are assigned slower modulation patterns to simulate large vessel behavior, while higher-frequency classes (Vessel 3 and 4) use faster modulations typical of smaller or more maneuverable ships~\cite{badiey2007frequency}.

\subsection{Statistical Texture Quantification via Bidirectional Temporal Entropy}
\label{sec:stat_score}

Statistical texture is quantified using a bidirectional entropy-based approach inspired by \cite{fraj2015temporal}, which evaluates how Shannon entropy accumulates over time in both forward and reverse directions. Let \( x(t) \) denote the signal over \( t \in [0, T_{\text{obs}}] \), sampled at rate \( f_s \), and segmented into overlapping frames of length \( L \) with hop size \( H \). For each frame \( i \), a normalized histogram with \( M \) bins yields a probability distribution \( \{p_{i,k}\} \), and entropy is computed as:

\begin{equation}
H_i = -\sum_{k=1}^{M} p_{i,k} \log_2 p_{i,k}.
\label{eqn:entropy}
\end{equation}

Entropy is accumulated for the original signal \( x(t) \) (forward) and its time-reversed form \( x_r(t) = x(T_{\text{obs}} - t) \) (reverse), producing curves \( H_d(i) \) and \( H_r(i) \). The area \( S \) between these captures directional asymmetry and is normalized by \( S_{\text{max}} = T_{\text{obs}} \cdot \log_2 M \). Relative entropy \( P_{\text{rel}} = H_{\text{final}} / \log_2 M \) and convergence proportion \( P_T \) are also computed. The final statistical texture score is:

\begin{equation}
\text{StaTS}_{\text{stat}} = 5 \cdot P_{\text{rel}} \cdot \left(1 - P_T \cdot \frac{S}{S_{\text{max}}} \right).
\label{eqn:stat}
\end{equation}

Higher scores indicate stronger statistical texture, marked by high entropy, fast convergence, and minimal asymmetry. This metric provides a compact and interpretable score that captures the degree of statistical texture by combining entropy magnitude, temporal symmetry, and convergence behavior.

\subsection{Structural Texture Quantification via Autocorrelation}
\label{sec:struct_score}
\begin{figure}[H]
    \centering
    \includegraphics[width=1\linewidth]{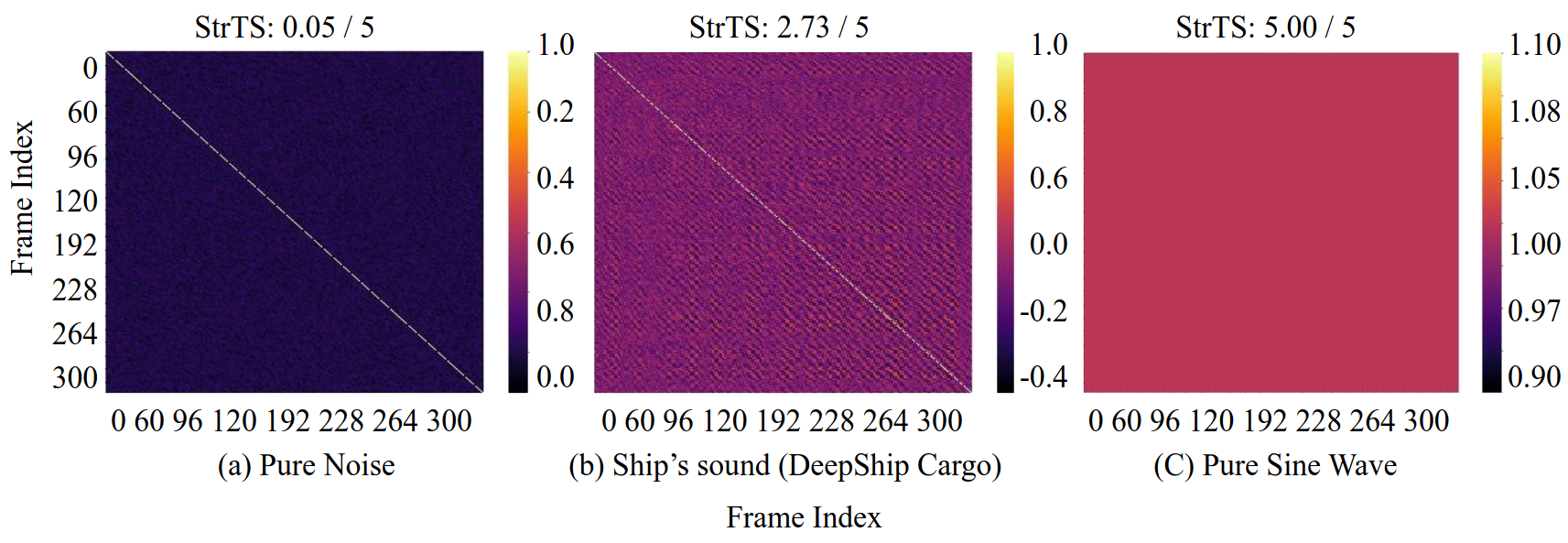}
    \vspace{-1em}
    \caption{
        Self-similarity matrices for three representative signals. 
        (a) White noise (StrTS = 0.05): lacks clear structure. 
        (b) Cargo ship signal (StrTS = 2.73): shows partial diagonal patterns. 
        (c) Pure sine wave (StrTS = 5.00): exhibits strong and evenly spaced diagonals.
    }
    \label{fig:self_similarity_examples}
\end{figure}

Structural texture is measured using a normalized autocorrelation-based score, capturing periodicity and harmonic regularity \cite{lee2008detecting}. The signal is first zero-meaned, and the autocorrelation is computed up to a lag \( \tau_{\text{max}} = 0.25 \cdot T_{\text{obs}} \). Only positive lags are considered, and the first significant peak is identified:

\begin{equation}
\tau^* = \arg\max_{\tau \in [1, \tau_{\text{max}}]} R(\tau),
\label{eqn:max_lag}
\end{equation}

The corresponding autocorrelation value at this lag, \( R(\tau^*) \), reflects the strength of the periodic structure and is scaled to produce the StrT in \ref{eqn:struct}:
\begin{equation}
\text{StrTS} = 5 \cdot R(\tau^*),
\label{eqn:struct}
\end{equation}

\noindent The score lies within the range \([0, 5]\), where higher values indicate stronger and more regular periodicity. For consistency, both StrTS and StaTS share this 0–5 scale, where lower scores indicate weak periodicity or less texture and higher scores reflect stronger structural or statistical patterns. 

In addition to the scalar score, self-similarity matrices offer a complementary visual representation of structural repetition. Clear, diagonal patterns correspond to higher StrTS scores. Figure~\ref{fig:self_similarity_examples} illustrates this with representative examples. The resulting score reflects structural periodicity in a normalized scale, making it useful for comparing a wide range of sonar signals and analyzing temporal regularity.

\section{Experimental Procedure}

\subsection{Dataset Description}
\begin{table}[ht]
\caption{Summary of datasets with the number of classes and total sample counts.}
\label{tab:dataset_summary}
\begin{center}
\small
\setlength{\tabcolsep}{10pt}
\renewcommand{\arraystretch}{1.2}
\begin{tabular}{|l|c|c|c|}
\hline
\textbf{Dataset} & \textbf{Type} & \textbf{Classes} & \textbf{Total Samples} \\
\hline
Synthetic-Structural   & Synthetic   & 4 & 40,000 \\ \hline
Synthetic-Statistical  & Synthetic   & 4 & 40,000 \\ \hline
Synthetic-Balanced     & Synthetic   & 4 & 40,000 \\ \hline
DeepShip~\cite{irfan2021deepship}     & Real-world  & 4 & 33,770 \\ \hline
VTUAD~\cite{domingos2022investigation} & Real-world  & 5 & 175,965 \\
\hline
\end{tabular}
\end{center}
\end{table}

Experiments are conducted on three synthetic and two real-world datasets to evaluate model performance under varying audio texture conditions. The real-world datasets DeepShip~\cite{irfan2021deepship} and Vessel Type Underwater Acoustic Data (VTUAD)~\cite{domingos2022investigation} offer diverse passive sonar recordings. VTUAD is the largest with 175{,}965 samples across five classes, while DeepShip provides 33{,}770 labeled samples across four classes. Table~\ref{tab:dataset_summary} summarizes the datasets by type, number of classes, and total sample count.

\subsection{Experimental Design}
\begin{table}[ht]
\caption{Comparison of structural (StrTS) and statistical (StaTS) texture scores along with classification accuracies of TDNN~\cite{peddinti2015time} and HLTDNN~\cite{ritu2023histogram} across five datasets. The last column shows the absolute accuracy improvement ($\Delta$Acc) of HLTDNN over TDNN. The best average accuracy is shown in bold.}
\label{tab:structural_dataset}
\begin{center}
\small % 10 pt text
\setlength{\tabcolsep}{8pt}
\renewcommand{\arraystretch}{1.2}
\begin{tabular}{|l|c|c|c|c|c|}
\hline
\textbf{Dataset} & \textbf{StrTS} & \textbf{StaTS} & \textbf{TDNN Acc. (\%)} & \textbf{HLTDNN Acc. (\%)} & \textbf{$\Delta$Acc. (\%)} \\
\hline
Synthetic-Structural    & 4.96 $\pm$ 0.03 & 3.52 $\pm$ 0.09 & 72.50 $\pm$ 3.38 & \textbf{99.67 $\pm$ 0.15} & +27.17 \\ \hline
Synthetic-Statistical   & 3.76 $\pm$ 0.35 & 4.93 $\pm$ 0.04 & 34.85 $\pm$ 1.04 & \textbf{55.05 $\pm$ 1.13} & +20.20 \\ \hline
Synthetic-Balanced      & 4.39 $\pm$ 0.21 & 4.33 $\pm$ 0.56 & 72.18 $\pm$ 3.21 & \textbf{99.08 $\pm$ 0.88} & +26.90 \\ \hline
DeepShip                & 3.16 $\pm$ 0.64 & 4.43 $\pm$ 0.55 & 50.49 $\pm$ 0.74 & \textbf{59.62 $\pm$ 1.69} & +8.06 \\ \hline
VTUAD                   & 4.45 $\pm$ 0.16 & 4.76 $\pm$ 0.32 & 71.92 $\pm$ 2.86 & \textbf{83.45 $\pm$ 0.57} & +10.53 \\ 
\hline
\end{tabular}
\end{center}
\end{table}

All signals are resampled to 32~kHz and segmented into five-second intervals, except VTUAD, which provides one-second clips with predefined train/validation/test splits. Log-mel spectrograms are computed using a Hanning window (size 1024, hop 320) with 1024 mel bins. No data augmentation is applied to preserve native texture characteristics.For all other datasets, a 70/10/20 train/validation/test split is used. TDNN serves as the baseline model, while HLTDNN incorporates a histogram-based feature layer to enhance statistical texture modeling. Both models are trained for 100 epochs using the Adagrad optimizer (learning rate \(1 \times 10^{-3}\), batch size 128). In addition to classification performance, each dataset is
analyzed using the proposed texture quantification metrics:
Structural Texture Score (StrTS) and Statistical Texture Score
(StaTS). These scores provide insight into the texture proper-
ties of each dataset and their influence on model performance.

\section{Results and Discussion}

\subsection{Synthetic and Real-World Datasets Evaluation}

Table~\ref{tab:structural_dataset} summarizes the structural (StrTS) and statistical (StaTS) texture scores, along with classification accuracies for TDNN and HLTDNN across five datasets. The final column reports the absolute accuracy improvement (\(\Delta\) Accuracy) of HLTDNN over TDNN. The results support the hypothesis that HLTDNN, with its histogram-based representation, is better suited for statistically rich signals. In the statistical texture dataset, where StaTS is highest (\(4.93 \pm 0.04\)), HLTDNN achieves a substantial +20.20\% improvement over TDNN, validating its ability to capture statistical variations more effectively. The structural texture dataset exhibits the highest StrTS (\(4.96 \pm 0.03\)) and lower StaTS. While TDNN performs reasonably well (\(72.50 \pm 3.38\)), HLTDNN still surpasses it with a +27.17\% gain, indicating that statistical modeling does not hinder its ability to capture structural patterns.

In the mixed statistical and structural texture dataset, HLTDNN achieves near-perfect accuracy (\(99.08 \pm 0.88\)) and shows a +26.90\% gain over TDNN, demonstrating adaptability in handling hybrid texture profiles. On the DeepShip dataset, which presents moderate statistical and lower structural texture complexity, HLTDNN achieves an +8.06\% improvement. Despite inherent variability in this real-world dataset, the performance gain reflects HLTDNN’s effectiveness in generalizing to practical scenarios. For the VTUAD dataset, which is the largest and exhibits high values in both StrTS and StaTS, HLTDNN delivers a strong +10.53\% improvement. This confirms the model’s scalability and its ability to learn from complex, large-scale real-world data.

\subsection{Pretraining Models Using Synthetic Datasets}

\begin{table}[ht]
\caption{Performance (\%) on DeepShip with pretraining using different synthetic datasets. Gains are reported relative to the baseline: TDNN = 50.49 ± 0.74, HLTDNN = 59.62 ± 1.69.}
\label{tab:pretrain_analysis}
\begin{center}
\small
\setlength{\tabcolsep}{6pt}
\renewcommand{\arraystretch}{1.2}
\begin{tabular}{|l|c|c|c|c|}
\hline
\multirow{2}{*}{\textbf{Pretraining Dataset}} & \multicolumn{2}{c|}{\textbf{TDNN}} & \multicolumn{2}{c|}{\textbf{HLTDNN}} \\
\cline{2-5}
 & Accuracy & Gain & Accuracy & Gain \\
\hline
Synthetic-Statistical              & 47.21 ± 1.89 & –3.28 & 53.21 ± 1.89 & –6.41 \\ \hline
Synthetic-Structural               & 50.63 ± 1.95 & +0.14 & 56.88 ± 1.22 & –2.74 \\ \hline
Synthetic Statistical and Structural & \textbf{52.50 ± 1.29} & \textbf{+2.01} & \textbf{62.11 ± 1.36} & \textbf{+2.49} \\
\hline
\end{tabular}
\end{center}
\end{table}

To evaluate the impact of synthetic pretraining, both models were pretrained on synthetic datasets and then fine-tuned on DeepShip. As shown in Table~\ref{tab:pretrain_analysis}, pretraining with the mixed texture dataset improved performance for both TDNN and HLTDNN, with a gain of +2.01\% and +2.49\%, respectively. In contrast, statistical-only or structural-only pretraining resulted in minimal or negative gains. This indicates that synthetic pretraining is particularly effective when the texture characteristics of the synthetic and real-world domains are well aligned. The mixed synthetic statistical \& structural dataset, which captures both statistical and structural properties, more closely reflects the texture complexity present in DeepShip, leading to better generalization after pretraining.

\section{Conclusion}
\label{sec:conclusion}

This work explored the role of structural and statistical texture in CNN-based passive sonar classification. Statistically rich signals, common in passive sonar due to stochastic propagation and interference, are often underrepresented in standard CNN processing. Synthetic datasets with controlled texture characteristics enabled systematic evaluation of model performance under varying texture complexities. Results show that while TDNN favors structural regularities, the texture-aware HLTDNN performs better on statistically textured signals, highlighting the value of modeling statistical texture. The proposed texture quantification metrics offer interpretability by linking signal properties to model behavior. Future work includes adapting foundation models like AudioPaLM~\cite{rubenstein2023audiopalm} and AudioGPT~\cite{huang2024audiogpt}, evaluating transformer-based models such as AST~\cite{gong2021ast}, integrating texture-aware regularization, using diffusion-based generation~\cite{kreukaudiogen}, and extending this framework to other audio domains.

% \acknowledgments % equivalent to \section*{ACKNOWLEDGMENTS}       
 
% This work was supported by the Under Secretary of Defense for Research and Engineering under Air Force Contract No. FA8702-15-D-0001 or FA8702-25-D-B001.

% \textbf{DISTRIBUTION STATEMENT A.} Approved for public release. Distribution is unlimited.

% \copyright~2025 Massachusetts Institute of Technology. Delivered to the U.S. Government with Unlimited Rights, as defined in DFARS Part 252.227-7013 or 7014 (Feb 2014).

% Any opinions, findings, conclusions, or recommendations expressed in this material are those of the authors and do not necessarily reflect the views of the Under Secretary of Defense for Research and Engineering.

% References
\bibliography{report} % bibliography data in report.bib
\bibliographystyle{spiebib} % makes bibtex use spiebib.bst

\end{document}